\documentclass[12pt]{article}

\def\pd{\partial_{\mu}}
\def\pa{\partial}
\def\pu{\partial^{\mu}}
\def\l{\lambda}
\def\m{\mu}
\def\n{\nu}
\def\e{\epsilon}
\def\s{\sigma}

\def\g{\gamma}

\def\a{\alpha}
\def\b{\beta}
\def\p{\phi}
\def\r{\rho}

\def\F{{\cal F}}

\def\be{\begin{equation}}
\def\ee{\end{equation}}
\def\ba{\begin{eqnarray}}
\def\ea{\end{eqnarray}}
\def\z{{ }^{*}}
\def\hn{\hat N}
\def\h{\hat}
\def\t{\tilde}
\def\f0{f_{0}}
\begin{document}

\begin{titlepage}   
\setcounter{page}{0}

\begin{flushright}

SINP-2002-11/695
\end{flushright}

\vskip 20pt

\begin{center}

{\large\bf  Near-horizon behavior
of string-loop-corrected dyonic black holes }
\vspace{2cm}

 {\bf Mikhail Z. Iofa \footnote{Skobeltsyn Institute of
Nuclear Physics, Moscow State University, Moscow 119992, Russia, e-mail:
iofa@theory.sinp.msu.ru}} 
\vskip 7pt
Skobeltsyn Institute of Nuclear Physics\\
 Moscow State University\\ 
 Moscow 119992, Russia
\vskip 7pt
\today
\end{center}

\vskip 20pt

\begin{abstract}
Dyonic black holes with string-loop corrections
 are studied in the near-horizon region.
In perturbative heterotic string theory compactified 
to four dimensions with $N=2$
supersymmetry, in the first order in string-loop expansion parameter, we
solve the system of Maxwell and Killing spinor equations for
dyonic black hole. At the horizon, the string-loop-corrected solution
displays restoration of spontaneously broken supersymmetry.
\end{abstract}
\vspace{1cm}
\vspace*{\fill}
\end{titlepage}

Since appearance of papers \cite{gi} , where it was noted that the
Robertson-Bertotti geometry is maximally supersymmetric, i.e.
does not break any of supersymmetries of $N=2,\,d=4$ supergravity, there
appeared a number of papers discussing this and related phenomena
in different systems \cite{kal,kapeet,k5,fk}.

In particular, it was made clear that if at special points
of space-time the metric of a system is asymptotic
to the Robertson-Bertotti metric, than at these points some partially
broken supersymmetries can restore.

This property of the metrics imply the universality properties
of black holes in supersymmetric theories: the entropy-area formula,
attractor properties of a system at the points of supersymmetry
restoration, etc. \cite{k5,fk,fks,bko,b5,b6,bls,4c,mo,bt1} and refs. therein.

In this letter we discuss these properties in the case
of the string-loop-corrected dyonic black hole. In the framework
of perturbative heterotic string theory compactified to  
$4D$ with $N=2$ supersymmetry, the prepotential of the effective
low-energy theory (so-called STU model) 
receives only one-string-loop correction from
string world sheets of torus topology \cite{afgnt,wikalu,hm}. 
String-loop expansion parameter
is  $\e =e^{\p_\infty}$, where
  $\p_\infty = \lim \p |_{r\rightarrow\infty}$.
Starting from the loop-corrected prepotential, in the first order in $\e$,
we solve Killing spinor equations (conditions for supersymmetry variations
of spinors to vanish). In the region of small $r$, near the horizon of the
black hole, we obtain expressions for the metric and moduli and verify
vanishing of certain objects implying restoration of supersymmetry.
  
$N=1$ supersymmetric $6D$ effective action of the heterotic string theory
compactified to
$4D$ on a two-torus $T^2$ yields $N=2$ locally supersymmetric theory.
Bosonic part of the $4D$ effective action in the Einstein frame is
\cite{sen}
\begin{equation}
\label{c1}
I_4 = \int d^4 x \sqrt{-g}\left[ R
-{1\over2}(\partial {\p } )^2 -{e^{-\p }\over4} \F(LML)\F +
\frac{a_1}{4\sqrt{-g}}\F L\z\F +\frac{1}{8}Tr (\partial ML\partial ML) \right ].
\end{equation}
On the other hand, the bosonic part of the action of a $N=2$ 
locally supersymmetric theory has the form \cite{dewit,andr,cafpr}
\begin{equation}
\label{c2}
I^{N=2}_4 = \int d^4 x \sqrt{-g}\left[\frac{1}{2} R + (\bar{N}_{IJ}
\F^{-I}\F^{-J}-
N_{IJ}\F^{+I}\F^{+J} ) + k_{i\bar{j}} \pd z^i \pu \bar{z}^j +\ldots\right] .
\end{equation}
Here $\F^{\pm }_{\m \n}=\frac{1}{2}(\F_{\m\n} \pm i\sqrt{-g}\z\F_{\m\n})$,
where $\z \F_{\m\n} ={1\over2}e_{\m\n\r\l}\F^{\r\l}$, and $e_{\m\n\r\l}$ is
the flat antisymmetric tensor, $e_{0123} =-1$.
The couplings $N_{IJ}$ are defined below,  $k_{i\bar{j}}$ is the K\"{a}hler
metric
$ k_{i\bar{j}}=\frac{\pa^2 K}{\pa z^i \pa \bar{z}^j }, $
where $K$ is the K\"{a}hler potential.
The tree-level moduli are 
\begin{eqnarray}
\frac{X^1 }{X^0 }=z^1 =iy_1 =i \left(e^{-\p } +ia_1 \right ), \nonumber \\
\frac{X^2 }{X^0 }=z^2 =iy_2 =i \left(e^{\g +\s} +ia_2 \right ),\nonumber \\
\frac{X^3 }{X^0 }=z^3 =iy_3 =i \left(e^{\g -\s} +ia_3 \right )
\label{c4}
\end{eqnarray}
Here $\p$ and $a_1$ are dilaton and axion, $a_2 =B_{12}$, other moduli 
are identified by comparison with the metric of the two-torus
\be
G_{mn}=e^{2\s}\left(\begin{array}{cc}
e^{2\g -2\s} +a_3^2 & -a_3 \\
-a_3 & 1
\end{array}\right)
\label{c5}
\ee
In the holomorphic section  which admits
introduction of the prepotential, the effective action (\ref{c1}) can be
identified with the action (\ref{c2}) having the prepotential of the STU 
model (for instance, \cite{wikalu})
\begin{equation}
F=-\frac{X^1 X^2 X^3}{X^0} -i\e {X^0}^2 h(-i\frac{X^2}{X^0},
-i\frac{X^3}{X^0})+\ldots.
\label{c6}
\end{equation}
At the one-loop level dilaton mixes with other moduli and
\be
y_1 = e^{-\p} -\e\frac{V}{2} +ia_1 .
\ee
The K\"{a}hler potential is given by 
\be
K=-\ln[(y_1 +\bar{y}_1 +\e V)(y_2 +\bar{y}_2)(y_3 +\bar{y}_3)],
\label{c7}
\ee
 and is invariant under symplectic transformations.
Here $V$ is the Green-Schwarz function $V$  \cite{wikalu,kou}
\be
V(y_2,\bar{y}_2,y_3,\bar{y}_3) = \frac{Re\,h -Re\,y_2 Re\,\pa_{y_2 }
 h -Re\,y_3 Re\,\pa_{y_3 } h}{Re\,y_2 \,Re\,y_3}.
\label{c8}
\ee
In the  section with the prepotential the gauge couplings are calculated using
the formula (for instance, \cite{dewit,andr,cafpr})
\be  
N_{IJ} = \bar{F}_{IJ} +2i \frac{(Im F_{IK}\, X^K ) (Im F_{JL}\, X^J )}
{(X^I\, Im F_{IJ}\,X^J )},
\label{c9}
\ee
where  $F_I =\pa_{X^I}F, F_{IJ}=\pa^2_{X^I X^J}F$, etc.

In the case of black hole solutions in which we are interested
in this paper, the
tree-level moduli $y_i$ are real. Since  the expressions of the
 first order in string
coupling  are calculated by substituting for the arguments the tree-level
moduli,  below, in cases where this does not lead to
confusion, we use the same notation $y_i$ for the real parts of the moduli.
Imaginary parts of the moduli corresponding to dyonic solution may appear in
the first order in string coupling.

In the first order in
string coupling constant, the gauge couplings  $N_{IJ}$ are \cite{mi} 
\ba
\label{c10}
N_{00}& =&iy^3\left(-1+\e\frac{n}{4y^3 }\right),\quad
N_{01}=-\e\frac{n+2v}{4y_1 }
-i\e a_1\frac{y_2 y_3}{y_1 },
 \nonumber \\
N_{02}&=&-\e\frac{n+2v-2y_2 h y +4y_2 h_2 }{4y_2 }
-i\e a_2\frac{y_1 y_3}{y_2 }, \nonumber \\
N_{03}&=&-\e\frac{n+2v+2y_3 h y +4y_3 h_3 }{4y_3 }
-i\e a_3\frac{y_1 y_2}{y_3 },\nonumber \\
N_{11} &=&-i\frac{y^3}{y_1^2 }\left(1+\e\frac{n}{4y^3 }\right),\quad
N_{12} =iy_3 \e\frac{2y_2 h y-n}{4y^3 } +\e a_3,\nonumber\\
N_{13}& =&iy_2 \e\frac{2y_3 h y-n}{4y^3 } +\e a_2,\quad
N_{23}= iy_1\e \frac{2yhy -4y_2 h_{23}y_3 -n}{4y^3} +\e a_1,\nonumber \\
N_{22}&=& -i\frac{y^3}{y_2^2 }\left(1-\e\frac{y_2 h_{23}y_3 }{y^3 }+
\e\frac{n}{4y^3 }\right),\quad
N_{33}= -i\frac{y^3}{y_3^2 }\left(1-\e\frac{y_3 h_{23}y_3 }{y^3 }+
\e\frac{n}{4y^3 }\right).
\ea
Here we introduced the notations: $y^3 = y_1 y_2 y_3,\,\, hy =h_a y_a =h_2
y_2
+h_3 y_3 $, $h_a =\pa_{y_a} h,\, h_{ab}=\pa_{y_a}\pa_{y_b}h$  and       
\be
\label{c11}
 v= h-y_a h_a, \qquad n= h- h_a y_a +y_a h_{ab} y_b, \qquad y_2 hy =y_2
h_{2a}y_a.
\ee

The field equations and the Bianchi identities for the gauge field strengths
are
\ba
\label{c12}
\pd \left( \sqrt{-g} Im \,G^{-\,\m\n }_I \right) =0 \nonumber \\
\pd \left (\sqrt {-g}Im\, \F^{-J\,\m\n} \right) =0
\ea
where $G^{-\,\m\n }_I=\bar{N}_{IJ}\F^{-J\,\m\n}$.
In sections which do not admit a prepotential (including that which
naturally appears in compactification of the heterotic string action), the
gauge
couplings are obtained by making a symplectic transformation of the
couplings  calculated in the section with the prepotential
\be
\hn=(C+D{N})(A+B{N})^{-1}.
\label{c13}
\ee

In the first order in string coupling,  transformation from the section
with the prepotential to that associated with the heterotic string
compactification is performed  by symplectic transformation with the matrices
\ba
A&=& diag (1,0,1,1)+\e (a_{ij}),\qquad B=diag (0,1,0,0)+\e (b_{ij}),
\nonumber \\
C&=& diag(0,1,0,0)+\e (c_{ij}),\qquad D=diag (1,0,1,1)+\e (d_{ij}),
\label{c14}
\ea
where $a,b,c$ and $d$ are constant symmetric  matrices. 
The form of the corrections to the tree-level matrices \cite{wikalu} is
constrained by the requirement that in the heterotic section
the loop corrections to the couplings are proportional to $\e e^\p $.
This results in symplectic transformation with $a=b=0$, $c$ is an arbitrary
symmetric matrix with $c_{1i}=0$ and the only non-zero element of the matrix
$d$ is $d_{11}$ \cite{mi}.

General dyonic solution can be obtained by solving the Killing spinor
equations which are conditions for the supersymmetry transformations of the
chiral gravitino $\psi_{\a\m }$ and gaugini $\l^{i\a}$ to vanish
(for instance, \cite{andr,stro,fre})
\ba
&{}&\delta \psi_{\a\m} =D_\m \e_\a - T^-_{\m\n}\g^\n \e_{\a\b}\e^\b =0, \\
\label{c18}
&{}&\delta \l^i_\a = i\g^\m \pd z^i \e^\a +
G^{-i}_{\m\n}\g^\m\g^\n\e^{\a\b}\e_\b =0,
\label{c19}
\ea
where
$$
D_\m \e_\a = (\pd -{1\over 4}w^{{\h{a}} {\h{b}}}_\m \g_{{\h{a}}} \g_{{\h{b}}} +
{i\over 2} Q_\m )\e_\a .
$$
Here $w_\m^{\h{a} \h{b}}$  and $Q_\m$ are the spin and K\"{a}hler
connections, and one introduces  symplectic invariants
\cite{andr,che,bls}
\ba
&{}&S_{\m\n}= X^I Im\,N_{IJ} \F^{-J}_{\m\n},\quad
T^-_{\m\n} =2ie^{K/2} S_{\m\n}, \\
&{}&G^{-i}_{\m\n} =-k^{i\bar{j}}\bar{f}^I_{\bar{j}} Im\,N_{IJ} \F^{-J}_{\m\n}.
\label{c17}
\ea
Here $k^{i\bar{j}}$ is the inverse K\"{a}hler metric, and
$$f^I_i = (\pa_i+\frac{1}{2}\pa_i K) e^{K/2}X^I .
$$
In the case of a stationary spherically-symmetric solutions with the metric 
\be
\label{c20}
 ds^2 =-e^{2U(r)} (dt + w_m (r) x^m )^2 + e^{-2U(r)} (dr^2 + r^2 d{\Omega}^2),
\ee
The $\m =0$ component of the gravitini  Killing spinor equation 
can be presented as
\be
\label{c21}
w_0^{-\h{a}\h{b}}\g_{\h{a}}\g_{\h{b}}\e_\a - 
e^U \,T^-_{0n}\g_{\h{n}}\e_{\a\b}\e^{\b} =0
\ee
Here the indices with hats refer to the tangent space basis.
Assuming the Ansatz for a constraint on the supersymmetry parameter
$\e^\a =\g_{\h{0}}\e^{\a\b}\e_\b $ (cf. \cite{fre,bls}), 
separating the coefficient at the
spinor structure and taking the real part
of the resulting equation,  we have
\be
\label{c21b}
\frac{1}{4}\pa_n e^{2U}-e^U\,Re\,T^-_{0n} =0,
\ee
where we used that $w_0^{\h{0}\h{b}}=\frac{1}{2}\pa_b e^{2U}$.
Substituting the Ansatz  in gaugini 
 Killing equation, we obtain 
\be
\label{c22a}
\left(ie^U \pa_n z^i -4G^{-i}_{0n} \right)
\g^{\h{0}} \g^{\h{n}}\e^{\a\b}\e_{\b} =0.
\ee 
Contracting Eq.(\ref{c22a}) with the functions $f^I_i$, we have
\be
if^I_i \pa_n z^i e^U + 
4\left({1\over 2}\F^{-I}_{0n}+ e^K\bar{X}^I S_{0n}\right)=0.
\label{c22}
\ee
Using the relations of special $N=2$ geometry,
Eqs.(\ref{c22}) can be recast in the form which contains 
$G^-_{I\,0n}$ and $\bar{F}_I $. Contracting Eq.(\ref{c22}) with $\bar{F}_I $
and using identities
$$\F^{-I} \bar{F}_I = G^-_{I} \bar{X}^I, \quad
\bar{F}_I f^I_i = \bar{X}^I g_{Ii},$$
where $g_{Ii} = (\pa_i+\frac{1}{2}\pa_i K) e^{K/2}F_I$, which follow from
definitions of $f^I_i$ and $G_{I\,\m\n}$, we have
$$\bar{X}^I g_{Ii}\pa_n  z^i + 4e^{-U}\left({1\over 2}G^-_{I\,0n}\bar{X}^I +
e^K \bar{F}_I \bar{X}^I S_{0n}\right) =0. $$
Removing the functions $\bar{X}^I$, we obtain the symmetric equation
\be
\label{c23}
 g_{Ii} \pa_n  z^i +
4e^{-U}\left({1\over 2}G_{-I\,0n}+ e^K \bar{F}_I S_{0n}\right)
=0.
\ee
Substituting the gravitino equation, Eqs.(\ref{c22}) and (\ref{c23}) are
presented as (cf. \cite{4c})
\ba
\label{c24}
 -2\F^{-I}_{0n} =i\left[e^U \pa_n (e^{K/2}X^I )- (e^{K/2}\bar{X}^I )\pa_n
e^U\right]+
2Im\,T^-_{0n} e^{K/2}\bar{X}^I -
Im (\pa_i K \pa_n y_i )e^{K/2 +U}  X^I ,\\
\label{c25}
 -2G^-_{I\,0n} =i\left[e^U \pa_n (e^{K/2}F_I )- (e^{K/2}\bar{F}_I )\pa_n e^U
\right]+2Im\,T^-_{0n} e^{K/2}\bar{F}_I -Im (\pa_i K \pa_n y_i )e^{K/2 +U} F_I.
\ea
Here we used the equality $ \pa_i K \pa_n z^i  =\frac{1}{2}\pa_n K
+i\,Im (\pa_i K \pa_n z^i ) $.  Eqs.(\ref{c24}) and (\ref{c25}) 
are not independent, but
one set can be obtained from the other. One can also take some equations
from the first set, and the remaining equations from the second.

Let us solve  Eqs.(\ref{c24}) and (\ref{c25}). 
We consider a tree-level dyonic black hole solution with vanishing axionic
parts, i.e. the tree-level moduli $y_i$  are real. Axions $a_i =Im\,y_i$
can appear at higher orders in $\e$.
The holomorphic section associated with the heterotic string
compactification is
\be
\label{c26}  
(\h{X}^I ,\h{F}_I ) =
 (1, y_2 y_3 , iy_2 , iy_3 ;\, -iy_1 y_2 y_3 , -iy_1 , y_1 y_3 -\e h_2 , 
y_1 y_2 -\e h_3 ).
\ee
The tree-level dyonic solution contains
two magnetic and two electric fields with strengths
\be
\z\h{\F}^0_{0r}=-\frac{P^0}{\sqrt{-g'}}, \qquad
\z\h{\F}^1_{0r}=-\frac{P^1}{\sqrt{-g'}}
\label{c29}
\ee   
and   
\be   
\h{\F}^2_{0r}=-\frac{Q_2}{\sqrt{-g'}Im \hn_{22}}, \qquad
\h{\F}^3_{0r}=-\frac{Q_3}{\sqrt{-g'}Im \hn_{33}},
\label{c30}
\ee   
where $\sqrt{-g' }= e^{-2U}r^2$ and the gauge couplings are
$$\hn_{00}=-iy_1 y_2 y_3 ,
\quad \hn_{11}=-i\frac{ y_1}{y_2 y_3},\quad
\hn_{22}=-i\frac{y_1 y_3}{ y_2},\quad \hn_{33}=-i\frac{y_1 y_2}{y_3}.
$$
 Since the tree-level moduli are real, $Im
(\pa_i K \pa_n z^i )=O(\e^2 ) $. 
Taking imaginary parts of Eqs.(\ref{c24}) for $I=0,1$ and Eqs.(\ref{c25})
for $I=2,3$, we obtain
\ba
\label{c27}
\sqrt{-g}\z\h{\F}^I_{0n} = e^{2U}\pa_n \left(e^{K/2-U}Re\,\h{X}^I\right)
+O(\e^2),\\
\label{c28} 
\sqrt{-g}\z\h{G}_{I\,0n} = e^{2U}\pa_n \left(e^{K/2-U}Re\,\h{F}_I\right)
+O(\e^2).
\ea
Substituting  (\ref{c29}) and (\ref{c30}) in the system (\ref{c27}) and 
(\ref{c28}), we obtain the equations
\be
\label{c31a}
-\frac{P^0}{r^2} = \left(e^{K/2-U}\right)', \quad
-\frac{P^1}{r^2} = \left(e^{K/2-U}y_2 y_3 \right)'
\ee
and
\be 
\label{c31b}
-\frac{Q_2}{r^2} = \left(e^{K/2-U}y_1 y_3 \right)',\quad
-\frac{Q_3}{r^2} = \left(e^{K/2-U}y_1 y_2 \right)'
\ee
with the tree-level solutions (cf. \cite{cvyo,cvyo1,bh2,bls,4c,bt1})
\ba
 e^{2U_0}=\frac{r^2}{(H^0 H^1 H_2 H_3)^{1/2}}, \quad
y_{1(0)} \equiv e^{-\p_0}\equiv\f0^{-1} =\left(\frac{H_2 H_3}{H^0
H^1}\right)^{1/2}, \\\nonumber
y_{2(0)} \equiv e^{\g_0 +\s_0}=\left(\frac{H^1 H_3}{H^0H_2}\right)^{1/2},
\quad y_{3(0)} \equiv e^{\g_0 -\s_0}=\left(\frac{H^1 H_2}{H^0 H_3}\right)^{1/2}.
\label{c32}
\ea
where
\be
H^0 =\sqrt{8}P^0 +a r, \quad H^1 =\sqrt{8}P^1 +a^{-1} r, 
\quad H_2 =\sqrt{8}Q_2 +b r, \quad H_3 =\sqrt{8}Q_3 +b^{-1} r.
\label{c33}
\ee
The constants are constrained by the requirement that 
solution is asymptotically flat. 

From the tree-level solutions it follows that 
 expressions  $\pa_n z^i$ are finite and the functions $G^{-i}_{\m\n}$ 
vanish
in the limit $r\rightarrow 0$. Introducing new variable, $\r =\frac{1}{r}$,
one obtains a conformally flat form of the metric
in the limit $\r\rightarrow\infty$. As follows from the above, in this limit
the derivatives of the moduli $\pa_\r z^i$ vanish.
Due to vanishing of the function $e^U$ in the limit $r\rightarrow 0$,
 all the terms in gravitini and gaugini equations
vanish separately and yield no restrictions on supersymmetry parameter
(cf. \cite{kal,kapeet,fk}), i.e. at this point supersymmetry is restored.   

In papers \cite{cvyo1} it was shown that in general case of string
tree-level dyonic black hole,
nonzero total electric and magnetic charges  yield  constraints, each
 preserving $1/2$ of the supersymmetry unbroken (so that general dyonic
configuration preserves $1/4$ of the supersymmetry unbroken).
This result is not in contradiction with the restoration of supersymmetry
at the horizon, because the constraints are obtained from Killing
spinor equations which contain the metric component $e^{2U}$. At the
horizon, the terms multiplied by the metric component $e^{2U}$ vanish, and
there appear no relations leading to constraints.

Let us turn to the loop-corrected dyonic solution. 
The functions
$\p, \g $ and $\s$ which enter the moduli  (\ref{c4}) and the function
$2U$ in
metric are split into the tree-level
parts $\p_{0} , \g_{0},\,\s_{0}, 2U_{0}$ and the parts of the first
order in string coupling: $\p =\p_{0} +\e\p_1,...,2U=2U_{0}+\e u_1$.
With the required accuracy, the K\"{a}hler potential is
\be
e^K = \frac{\f0 e^{-2\g_0 }}{8}\left[1 +\e\left(\p_1 -2\g_1 \right)\right].
\label{c36}
\ee
In the first order in string coupling, taking into account
that the terms of the main order in string coupling cancel,  
Eqs.(\ref{c27}) with $I=0,1$ yield
\ba
\label{c37}
\left(\p_1 -u_1 -2\g_1 
\right)' +\frac{1}{2}(\p_0 -2U_0 -2\g_0 )' \left(\p_1 -u_1 -2\g_1 \right)=0, 
\\\nonumber
\left(\p_1 -u_1 +2\g_1 \right)' +\frac{1}{2}(\p_0 -2U_0 +2\g_0 )' 
\left(\p_1 -u_1 +2\g_1  \right)=0.
\ea
From this system we find that in the limit $r\rightarrow\ 0$ both functions
$\g_1$  and $ \p_1 -u_1$ vanish as $O(r)$. 
Substituting the loop-corrected expressions $\h{F}_2 =y_1 y_3 -\e h_2$ and
$\h{F}_3 =y_1 y_2 -\e h_3$ and introducing  $L_2 =\e h_2 y_2 e^{-2\g_0}$
and  $L_3 =\e h_3 y_3 e^{-2\g_0}$, we
reduce the second pair of equations (\ref{c28}) to the form
\ba
\label{c39}
\left(\p_1 +u_1  +2\s_1 +(V+2L_2 ) \f0\right)' -
\frac{1}{2}\left(\s_0 +U_0 +\frac{\p_0}{2}\right)'
\left(\p_1 +u_1  +2\s_1+(V +2L_2 )\f0\right)=0,
\\\nonumber
\left(\p_1 +u_1 -2\s_1 +(V+2L_3 )\f0\right)' -
\frac{1}{2}\left(-\s_0 +U_0 +\frac{\p_0}{2}\right)'
\left(\p_1 +u_1 -2\s_1 +(V+2L_3 ) \f0\right)=0.
\ea   
From these equations, in the limit $r\rightarrow 0$ we find 
\be
\label{c40}
\p_1 =u_1 = -H\f0(0)=
-\frac{h(y_{2(0)},y_{3(0)})}{2y_{1(0)} y_{2(0)} y_{3(0)}}\biggr|_{r=0}=
\frac{h(y_{2(0)},y_{3(0)})}{2}\biggr|_{r=0}\left(\frac{{P^0}^3}{P^1 Q_2 
Q_3}\right)^{1/2}
\ee
and
\be
\label{c41}
\s_1 =\frac{1}{2} (L_3 -L_2 )\biggr|_{r=0}\left(\frac{P^0 P^1}{Q_2 
Q_3}\right)^{1/2} +O(r).
\ee
Here we introduced
\be
\label{c40a}
H=\frac{V +L }{2}\biggr|_{r=0}  =\frac{h(y_{2(0)},y_{3(0)})}{2 y_{2(0)} 
y_{3(0)}}\biggr|_{r=0}, 
\qquad L=\frac{L_2 +L_3}{2}.
\ee
Let us consider solution of the gaugini Killing spinor equations written in an 
alternative form (\ref{c22}). At the tree level, substituting
explicit expressions for the field strengths (\ref{c29}) and  
(\ref{c30}) and solutions for the moduli (\ref{c32}), we find that in the limit
$r\rightarrow 0$ all the combinations $\left({1\over 2}\h{\F}^{-I}_{0n}+
e^K \bar{\h{X}}^I S_{0n}\right)$ vanish implying that $\pa_n
y_i|_{r\rightarrow\ 0} $, are finite and
$G^i_{\m\n}$ vanish.
Vanishing of of the tensor $G^i_{\m\n}$ together with Bertotti-Robinson
form of the metric near the point $r=0$ ensure vanishing of the Weyl tensor
and conformal invariance of the theory in this region.

Keeping the terms of the
main and the first orders in string coupling, the Maxwell equations
$$\pd (\sqrt{-g}\, Im \hn_{IJ} \h\F^J +Re \hn_{IJ}\z \h\F^J )^{\m\n} =0.
$$
written in the heterotic holomorphic section  are
\ba
 \pa_r [\sqrt{-g} ( Im \hn_{00} \h\F^0 +Im \hn_{02} \h\F^2 +
Im \hn_{03} \h\F^3 ) + Re \hn_{00}\z\h\F^0 +
Re \hn_{01}\z\h\F^1 ]^{0r} =0 
\nonumber\\
 \pa_r [\sqrt{-g} ( Im \hn_{11} \h\F^1 +Im \hn_{12} \h\F^2 +
Im \hn_{13} \h\F^3 ) + Re \hn_{10}\z\h\F^0 + Re \hn_{11}\z\h\F^1 ]^{0r} =0
\label{c42}
\ea
\ba
\quad \pa_r [\sqrt{-g}(Im \hn_{22} \h\F^2 +Im \hn_{23} \h\F^3 ) +
Re \hn_{20}\z\h\F^0 + Re \hn_{21}\z\h\F^1  ]^{0r} =0 \nonumber\\
\quad \pa_r [\sqrt{-g}(Im \hn_{33} \h\F^2  +Im \hn_{32} \h\F^2 )+
Re \hn_{30}\z\h\F^0 + Re \hn_{31}\z\h\F^1  ]^{0r} =0 \label{c42a}.
\ea
Only the diagonal gauge couplings $\hn_{II}$ contain the parts of the main order
in string coupling. The field strengths $\h\F^{0,1}_{0r}$,
absent at the tree level, are of the first order in string coupling.
Solving the Maxwell equations  in the holomorphic
section associated with the heterotic string compactification 
with the gauge
couplings $\hn_{IJ}$ calculated from the couplings  (\ref{c9}) by using 
the transformations (\ref{c13}), (\ref{c14}), we obtain 
\ba
\h\F^{0\,0r}& =&\frac{q_0  -a_1\,P^1 -a_a\,Q_a}
{\sqrt{-g'}\,Im \hn_{00}},
\label{c43} \\
\h\F^{1\,0r}& =&\frac{q_1  -a_1\,P^0+a_3 Q_2
/y_3^2 +a_2 Q_3/y_2^2 } {\sqrt{-g'}\,Im \hn_{11}}, 
\label{c44}\\
\h\F^{2\,0r}&=&\frac{ Q_2 -Re \hn_{20}\,P^0
-Re \hn_{21}\,P^1 -\frac{Im N_{23}}{Im N_{33}}
\,Q_3 }{\sqrt{-g'}Im \hn_{22} },
\label{c45} \\
\h\F^{3\,0r}&=&\frac{Q_3 -Re \hn_{30}\, P^0
-Re \hn_{31}\, P^1 -\frac{Im N_{32}}{Im N_{22}}\,Q_2 }{\sqrt{-g'}Im N_{33} } 
\label{c46}.
\ea
Substituting the field strengths (\ref{c43})-(\ref{c46}), we have \cite{mi}
\ba
\label{c47}
&{}& S_{0n}=\{[P^0 (Im\, N_{00} +y_i Re N_{i0}  )  -P^1
y_1 -( Q_2 y_2 + Q_3 y_3 )]\\
&{}&-i\e [P^0 (a_1 y_2 y_3 +a_2 y_1 y_3 + a_3 y_1 y_2 )+a_a Q_a
 +a_1 P^1 - q_0   -  q_1  y_2 y_3 )]\}
\frac{i}{2}e^{2U}\frac{x^n }{r^3 }\nonumber
\ea
Because Killing spinor equations are linear in derivatives of the moduli, 
in the
first order in string coupling the equations for the real and imaginary
parts of the moduli decouple. To obtain the loop corrections to the dilaton
$\p$ and the metric component $e^{2U}$ as well as to the metric of the
two-torus $G_{mn}$ it is sufficient to solve the  Killing spinor equations
for the real parts of the moduli. At this stage, the imaginary parts
of the moduli can be neglected. 

The combinations $f^I_i \pa_n z^i$ which enter  Eqs.(\ref{c22}) are
\be
\label{c48}
f^0_i \pa_n z^ =\frac{1}{2\sqrt{8}}e^{1/2(\p_0 -2\g_0 )}({\p}'_0 -2{\g}'_0 )
\left[1 +\e\frac{{\p}'_1 -2{\g}'_1}{{\p}'_0 -2{\g}'_0 } +
\frac{\e}{2}\left({\p}_1 -2{\g}_1 \right)\right]\frac{x^n}{r}, 
\ee
and
\be
\label{c49}
f^1_i \pa_n z^ =\frac{1}{2\sqrt{8}}e^{1/2(\p_0 +2\g_0 )}({\p}'_0 +2{\g}'_0 )
\left[1 +\e\frac{{\p}'_1 +2{\g}'_1}
{{\p}'_0 +2{\g}'_0 } +\frac{\e}{2}\left({\p}_1  
+2{\g}_1 \right)\right]\frac{x^n}{r}.
\ee
Other combinations are $f^i_n \pa_n z^i = iy_i (f^0_n + e^{K/2}\pa_n y_a 
)$, where $i=1,2,3$.
The expression $e^K S_{0n}$   in Eqs.(\ref{c22}) can be presented as
\footnote{More exactly, here we consider gaugini Killing spinor 
equations for the
real parts of the moduli, i.e. for $S_{0n}$ 
 we take the first (real) term in curly brackets in (\ref{c47}). 
Imaginary parts of the moduli will be discussed
separately. }  
\be 
\label{c50}
e^K S_{0n}=\left(1-\e {V\f0\over2}\right)(S_0
+\e S_1)\frac{i}{16\sqrt{-g'}}\frac{x^n}{r},
\ee
where
\be
\label{c51}
S_0 =P^0 +P^1 e^{-2\g_0} +Q_2 e^{\p_0 -\g_0 +\s_0 }
+Q_3 e^{\p_0 -\g_0 -\s_0 } 
\ee
and
\ba
\label{c52}
&{}&S_1 =P^0 2(V+L)\f0 +P^1 e^{-2\g_0} (-2\g_1 )\\\nonumber
&{}&+ Q_2 e^{\p_0 -\g_0 +\s_0  }\left(\p_1 -\g_1 +\s_1 +\frac{V\f0}{2}\right )
+Q_3 e^{\p_0 -\g_0 -\s_0 } \left(\p_1 -\g_1 -\s_1 + \frac{V\f0}{2}\right  ).
\ea
Substituting the above expressions, we
obtain the gaugini Killing equations (\ref{c22}) with $I=0,1$ in the form
\ba
\label{c53}
\frac{{\p}'_1 -2{\g}'_1}{{\p}'_0 -2{\g}'_0 } +\frac{1}{2}\left({\p}_1
-2{\g}_1 -u_1 \right) -\frac{S_1 -S_0 \frac{V\f0}{2}}{4P^0 -S_0} =0,\\\nonumber
\frac{{\p}'_1 +2{\g}'_1}{{\p}'_0 +2{\g}'_0 } +\frac{1}{2}\left({\p}_1
+2{\g}_1 -u_1\right) -\frac{S_1
-S_0 \frac{V\f0}{2}}{4P^1 H^0 /H^1  -S_0} =0.
\ea
At small $r$ we have
\be
\label{c54}
S_0 =4P^0 +2P^0 r({\p }'_0 -2{\g }'_0 )(0) +O(r^2 )
\ee
and
\be
\label{c55}
S_1 = 2 P^0(\p_1 -2\g_1 +H\f0 )(0) +O(r).
\ee
  Eqs.(\ref{c53}) split into the following system
\ba
\label{c56}
{\g}'_1 +2{\g}'_0 (\p_1 -u_1 ) +2{\p}'_0 {\g}_1= 0, \\\nonumber
{\p}'_1+\frac{{\p}'_0}{2}( {\p}_1 -{u}_1 ) +2 {\g }'_0 {\g }_1
- \frac{S_1 -S_0 \frac{V\f0}{2}}{4P^0 -S_0} =0
\ea
Substituting (\ref{c54}) and (\ref{c55}) 
and noting that the functions $\p_1$ and $u_1$ are
finite at the origin, and $\g_1$ is of order $O(r)$, at small $r$ 
we reduce the second equation (\ref{c56}) to the form
\be
\label{c56a}
{\p}'_1- \frac{{\p}_1}{r} -\frac{H(0)\f0}{r} =0
\ee
Similar transformations of the gravitini equation yield
\be
\label{c57}
\frac{{u}'_1}{2U'_0} -\frac{u_1}{2}  +\frac{1}{2}({\p}_1
-2{\g}_1 -u_1) - \frac{S_1 -S_0 \frac{V\f0}{2}}{4P^0 -S_0}=0,
\ee
 At small $r\,\,2U'_0 =\frac{2}{r} +O(1)$, and the
gravitini Killing equation takes the same form as (\ref{c56a})
\be
\label{c58}
{u}'_1 -\frac{u_1}{r}  -\frac{H(0)\f0}{r} =0.
\ee
Solving  Eqs.(\ref{c56a}) and (\ref{c58}), we 
reproduce the solution  (\ref{c40}). 
In the same way we obtain the above solutions for $\g_1$ and $\s_1$.

Let us consider the equations for imaginary parts $a_i$ of the
moduli $y_i =Re\, y_i +i a_i$. 
The functions $a_i$ are zero at the tree level, and can appear in the first 
order in string coupling. Below, in the equations for $a_i$, 
the moduli $Re\,y_i$ appear
only in the main order, and  again we keep notations $y_i$ for the
real parts of the moduli.
Contracting gaugini Killing  Eq.(\ref{c22a}) with the
metric $k_{i\bar{j}}$, introducing  $T_{I0r}\equiv Im\,
N_{IJ} \F^{-J}_{0r}$ and separating the imaginary part
of the resulting equation, we have
\be
\label{c59}
k_{\bar{j}i} a'_i +4e^{-U} Im\, (\bar{f}^I_{\bar{j}} \,T_I )=0.
\ee
Introducing   $\t{q}_1 = q_1 e^{2\g_0}$, 
($y_{2(0)} y_{3(0)} =e^{2\g_0}$, see (\ref{c32}))
with the required accuracy,  Eqs.(\ref{c59}) are written as
\ba
\label{c60}
 a'_1 +\frac{4e^{U +K/2} y_1 }{r^2 }
\left[{q}_0 +\t{q}_1   -a_1 P^1 -a_a Q_a +
P^0 (-a_1 y_2 y_3 + a_2 y_1 y_3  +a_3 y_1 y_2 )\right]=0,\nonumber \\
 a'_2 +\frac{4e^{U+ K/2} y_2 }{r^2 }
\left[{q}_0 -\t{q}_1  -a_1 P^1 -a_a Q_a +
\h{P}^0 (+a_1 y_2 y_3 - a_2 y_1 y_3  +a_3 y_1 y_2 )\right]=0,\nonumber \\
 a'_3 +\frac{4e^{U+ K/2} y_3 }{r^2 }
\left[{q}_0 -\t{q}_1  - a_1 P^1 -a_a Q_a +
P^0 (+a_1 y_2 y_3 + a_2 y_1 y_3  -a_3 y_1 y_2 )\right]=0.
\ea  
Here all the functions $y_i, \,U$ and $K$ are taken in the main order in
string coupling. Let us introduce the functions $b_i =
a_i /y_i$. All  the coefficients
at the functions $b_i$ in Eqs.(\ref{c60}) rewritten through the functions $b_i$ 
\vspace{.3mm}
in the limit $r=0$ coincide and are equal 
to $\left(\frac{P^1 Q_2 Q_3}{P^0}\right)^{1/2}\equiv \a$, and the expression
$\frac{4e^{U +K/2} }{r^2 }$ 
\vspace{.2mm}
at small $r$ is $\frac{\sqrt{2}}{r\a}$.  
In the limit of small $r$, we obtain the system (\ref{c60}) 
 in the 
\vspace{.3mm}
form
\ba
\label{c61}
 {b'}_1 + b_1 \frac{{y'}_1}{y_1} + \frac{\sqrt{2}}{r\a}[{q}_0 +\t{q}_1 -
2\a b_1 +O(r)]=0, \\\nonumber
 {b'}_2 + b_2 \frac{{y'}_2}{y_2} + \frac{\sqrt{2}}{r\a}[{q}_0 -\t{q}_1 -
2\a b_2  +O(r)]=0, \\\nonumber
 {b'}_3 + b_3 \frac{{y'}_3}{y_3} + \frac{\sqrt{2}}{r\a}[{q}_0 -\t{q}_1 -
2\a b_3  +O(r)]=0.
\ea
with a solution
\ba       
\label{c62}
&{}&b_1 =\frac{{q}_0 +\t{q}_1 }{2\a } + c_1 r +O(r^2 ),
\\\nonumber
&{}&b_a =\frac{{q}_0 -\t{q}_1 }{2\a } + c_a r +O(r^2 ),
\quad a=2,3.
\ea
where  $c_i$ are arbitrary constants. 

Using the above asymptotics of the axions, we can consider
stationarity properties of the loop-corrected solution.
Taking the imaginary part of the gravitini spinor Killing equation 
(\ref{c21}), we have
\ba
\label{c63}
\frac{1}{2}\z w^{\h{0}\h{n}}_0 \g^{\h{0}}\g^{\h{n}}\e_\a + e^U
Im\, T^-_{0n}\g^{\h{0}}\g^{\h{n}}\e_{\a\b}\e^{\b} =0,
\ea 
where $\z w^{\h{0}\h{n}}_0 =\frac{1}{2}e^{\h{n}\h{p}\h{q}}(\pa_p w_q
-\pa_q w_p )$ and $Im\, T^-_{0n}=2 Re\, e^{K/2} S_{0n}$ with $S_{0n}$ from
(\ref{c47}). In the limit of small $r $ we have $Im \,T =O(r^0)$ 
and 
$e^U Im\, T = O(r)$ implying vanishing of the functions $w_n$ in this 
limit and stationarity of solution. Also, suitably adjusting the
free constants in (\ref{c62}),  the asymptotic
(physical) charges of the electric fields $\h{\F}^{0}$ and $\h{\F}^{1}$
can be made vanishing. 
 
To conclude, general loop-corrected dilatonic black holes with four charges, 
as the tree-level
solutions, at the horizon display restoration of $N=2$ supersymmetry. 
The metric in the near-horizon region becomes of the
Robinson-Bertotti type. This result is natural, provided perturbation
theory does not violate supersymmetry.

\vspace{.5cm}
{\large \bf Acknowledgments}

This work was partially supported by the RFFR grant No 02-02-16444.

\end{document}